\documentclass[a4paper,20pt]{article}
    \usepackage[T1]{fontenc}
    \usepackage{graphicx}
    \usepackage{epsfig}
    \usepackage{amsmath}
    \usepackage{amsfonts}
    \renewcommand{\abstract}{}
\begin{document}
\def\eq#1{Eq.\hspace{1mm}(\ref{#1})}
\def\fig#1{Fig.\hspace{1mm}\ref{#1}}
\def\tab#1{Table.\hspace{1mm}\ref{#1}}
\makeatletter
\renewcommand{\@oddhead}{\textit{YSC'14 Proceedings of Contributed Papers} \hfil \textit{M. Jarosik, S. Starzy{\'n}ski, et al.}}
\renewcommand{\@evenfoot}{\hfil \thepage \hfil}
\renewcommand{\@oddfoot}{\hfil \thepage \hfil}
\fontsize{11}{11} \selectfont

\title{The First Steps of Radio Astronomy in Cz\k{e}stochowa}
\author{\textsl{M. Jarosik$^{1}$, S. Starzy{\'n}ski$^{1}$, M. Szcz\k{e}{\'s}niak$^{2}$, R. Szcz\k{e}{\'s}niak$^{1}$, A. Ceglarek$^{1}$}}
\date{}
\maketitle
\begin{center} {\small $^{1}$Institute of  Physics, Cz\k{e}stochowa University of Technology, Al. Armii Krajowej 19,  42-200 Cz\k{e}stochowa,  Poland \\
szczesni@mim.pcz.czest.pl\\
\small $^{2}$Ul. S{\l}oneczna 44a, 34-700 Rabka,  Poland \\
kms.rabka@neostrada.pl\\
}
\end{center}

\begin{abstract}
In the paper, technical documentation and the principle of operation
is presented. "KLAUDIA" radio telescope was built in Rabka in 2007
and it is used to receive secondary radio waves, emitted by the
Earth's ionosphere at frequency of 40 kHz.
\end{abstract}

\section*{Introduction}
\indent \indent Construction of the radio telescope "KLAUDIA" was
finished in 2007 \cite{hutnik, hutnik2}. The radio telescope is
placed in Rabka - small resort in southern Poland. Low level of
noise was a deciding factor when looking for a useful location;
Rabka is a mountain resort situated far from big cities. Building of
the radio telescope went on for three years.

This publication was divided into three fundamental chapters. In the
first chapter the results of modal and strength analysis of the new
radio telescope are presented; numerical analysis was made by finite
elements method  \cite{mes}. Radio telescope was provided with
custom made VLF 40 kHz receiver manufactured in the USA \cite{jeff}.
The receiver can detect the radio waves radiation of the Earths
ionosphere which frequency is 40 kHz; construction of the VLF
receiver was described in detail in the second chapter.

In the third chapter the initial measurements of the ionosphere
radio waves is presented. It was recorded soon after activation of
the radio telescope.

"KLAUDIA" radio telescope will be used to constantly observe the
secondary radiation of the Earth ionosphere; also, it will be
perfect didactic base for people interested in radioastronomy.

\section*{Model of the radio telescope construction - strength and modal ana-\\lysis}
\indent \indent Fig.1 presents main elements of the radio telescope
steel construction. The model of the radio telescope was made in
AutoCAD. Fig.1 (a) presents main dish; main dish is aluminium sheet;
diameter of the focusing cup is 1.8 m. Fig.1 (b) and (c) adequately
present mast of the radio telescope and one of the steel shids which
all construction is based on.

On top of the model of the radio telescope a regular mesh has been
superimposed and it was made subject to modal and strength analysis
proving that supporting structure of the radio telescope is very
tough; it slightly deformed while experiencing significant strain
greater than 500 N. On the other hand, the modal analysis has shown
that frequency of the first transverse mode is equal to 3.012 Hz.
Vibration at this frequency produced by wind could lead to
appearance of the mechanical resonance and the radio telescope could
be damaged. Vibrations of the construction and their direction are
shown in Fig.2.

To eliminate the vibrations at frequency 3.012 Hz steel beams were
put between horizontal shids. Once more, modal analysis shows
minimal vibration of the construction.

Radio telescope was constructed based on results strength and modal
analyse. Fig. 3 shows the finished construction placed in Rabka.

\section*{Construction of the VLF 40 kHz receiver}
\indent \indent The main electronic assembly of the radio telescope
is a very sensitive, tuned radio frequency receiver. This receiver
is divided into few modules; the first module is the preamp with
antenna feed and includes a voltage regulator Zener diode. The first
stage also includes a very sharp bandpass filter resonant at 40 kHz
with bandwidth of  a few kHz. For the maximum gains adjust the two
coils which is visible in Fig.4 (a). The preamp also includes two
active devices designed with static electricity protection provided
by two diodes. The backend comprises of two additional gain stages
and another bandpass filter resonant at 40 kHz. Provision for RF
gain adjustment has been made by adding pot between stages. (Fig.4
(b)). Moreover, it serves a protection from overdriving of the
receiver. Operational amplifier provides DC gain for the detected
signal which is then sent to the integrator board. The integrator
which is based on RC is set to 10 second. Finally the signal is
sampled by a/d converter and sent via RS-232 (Fig.4 (d)) to PC
computer's COM port. The receiver is placed inside a metal cover
which shields from noise. The receiver is powered by a +12 V
standard stabilized DC supply. Electronic assembly consumes less
power than a power LED. Total power consumed by the receiver is less
than 1 W.

\section*{Measurement of the Earths ionosphere radio waves radiation}
\indent \indent "KLAUDIA" radio telescope is oriented to investigate
Sun activity by receiving secondary radio waves of the Earth
ionosphere \cite{radioastronomia}, \cite{pal}; radio telescope is
receiving electromagnetic waves from VLF (Very Low Frequency) band
(40 kHz). The VLF receiver connects with a PC via the COM port. To
record a signal on PC computer the program "SpectraCyber" is used
\cite{cd}. This software enables setting sampling frequency of the
signal, online view the course of signal and saving data to a file.
Example of measurements of the Earth's ionosphere radio waves is
presented in Fig.5; The vertical axis presents normalized voltage
outputs of the VLF receiver 40 kHz, horizontal axis represents time.
Measurements took 27 hours; they were started on 15.03.2007 at
midnight.

On the Fig.5 a maximum is visible which corresponds to the daily
radiation of the Earth ionosphere and two minima for the night. From
the graph results that radioastronomical day has got 11 hours and 36
minutes.

\section*{Conclusions}
\indent \indent Construction and principle of  operating the
"KLAUDIA" radio telescope was presented in the paper. The results of
the modal and strength analyse prove that construction of the radio
telescope is designed correctly: construction is strong enough and
it has low propagation constant of vibration.

Included example measurement of the Earth's ionosphere radio waves
allows us to believe that the radio telescope  works correctly. In
the future development and modernization of the radio telescope is
planned; we are going to equip our radio telescope with 4-meters
main dish and assemble a multi-channel radio wave receiver.

\section*{Acknowledgements}
\indent \indent Authors are grateful to Dr. Jaros\l{}aw Solecki for
priceless help in realization of this project and Dr.~Bogdan
Wszo\l{}ek for supervising and consulting.

\newpage

\textbf{Figure 1.} Main construction elements of the "KLAUDIA" radio
telescope: a) dish, b) mast c) steel shid.\vspace{10ex}

\textbf{Figure 2.} Visualisation of the vibrations of the "KLAUDIA"
radio telescope construction. The frequency of the vibrations is
3.012 Hz and they are characterized by great amplitude.\vspace{10ex}

\textbf{Figure 3.} Construction prosess of the "KLAUDIA" radio
telescope built in Rabka.\vspace{10ex}

\textbf{Figure 4.} Photography presents choosen electronical
elements of the VLF 40 kHz receiver: a) coils, b) pot, c)  antenna
socket, d) signal otput.\vspace{10ex}

\textbf{Figure 5.} Normalized voltage outputs of the VLF receiver 40
kHz in time function.\vspace{10ex}

Figures are available on YSC home page
(http://ysc.kiev.ua/abs/proc14$\_$16.pdf).

\end{document}